\documentclass[prd,twocolumn,showpacs,preprintnumbers,amsmath,amssymb,superscriptaddress,floatfix,nofootinbib]{revtex4}

\usepackage{graphicx}
\usepackage{epstopdf}
\usepackage{bm}
\usepackage{amsmath}
\usepackage{amsfonts}
\usepackage{amssymb}
\usepackage{color}
\usepackage{multirow}
\usepackage{footnote}

\usepackage[colorlinks, citecolor=blue,anchorcolor=red,menucolor=red,linkcolor=red,filecolor=red,runcolor=red,urlcolor=blue,frenchlinks=red]{hyperref}
\begin{document}
\title{Coupled Channel Effects for the Charmed-Strange Mesons}
\author{Wei Hao}
\email{haowei2020@itp.ac.cn}
\affiliation{CAS Key Laboratory of Theoretical Physics, Institute of Theoretical Physics, Chinese Academy of Sciences, Beijing 100190,China}
\affiliation{School of Physical Sciences, University of Chinese Academy of Sciences (UCAS), Beijing 100049, China}

\author{Yu Lu}
\email{ylu@ucas.ac.cn}
\affiliation{School of Physical Sciences, University of Chinese Academy of Sciences (UCAS), Beijing 100049, China}

\author{Bing-Song Zou}
\email{zoubs@itp.ac.cn}
\affiliation{CAS Key Laboratory of Theoretical Physics, Institute of Theoretical Physics, Chinese Academy of Sciences, Beijing 100190,China}
\affiliation{School of Physical Sciences, University of Chinese Academy of Sciences (UCAS), Beijing 100049, China}
\affiliation{School of Physics and Electronics, Central South University, Changsha 410083, China}

\begin{abstract}
We make a systematic calculation of the spectra and hadronic decays of the $D_s$ system in a coupled channel framework, where the unquenched effects are induced by the $^3P_0$ model.
In the calculation, the wave functions are obtained by using a nonrelativistic potential model and are handled precisely with Gaussian Expansion Method. 
Even though the fitting mainly focuses on the spectrum, 
our model agrees well with the experiments on both the spectra and the hadronic decays,
suggesting that the coupled channel effect could result in a reasonable and coherent description of the $D_s$ mesons.
Based on the calculation, we give a detailed analysis on various aspects of the excited states, especially $D_s(2317), D_s(2460), D_s(2536), D_s(2860), D_s(3040)$.
We also predict that $D_s(2{}^3P_0)$ should be a $D^*K^*$ dominant molecule with mass $2894$~MeV, which is only $5$~MeV below the $D^*K^*$ threshold.

\end{abstract}

\maketitle

\section{Introduction}

Back in 1964, in order to build a schematic model of baryons and mesons, 
Gell-Mann \cite{Gell-Mann:1964ewy} and independently Zweig \cite{Zweig:1964ruk} proposed a fundamental component now called quarks to be the building blocks of hadrons.
In this model, baryons and mesons are compounds of three quarks and quark-antiquark pairs, respectively.
This breakthrough lays the playground for the development of Quantum chromodynamics (QCD), 
which is widely believed to be the fundamental theory of strong interactions.
However, due to the asymptotic freedom of the QCD, the coupling constant becomes comparable to one around $\sim 300$~MeV,
making the perturbation calculation approach infeasible.

Two general frameworks at the quark level have been proposed to circumvent this difficulty.
One way is the lattice QCD, where the space-time is discretized and the QCD is simulated on supercomputers.
Another way belongs to various QCD-inspired phenomenological models.

Within the phenomenological model approach, and an elaborate treatment of the relativistic effect from the gluon,
Godfrey and Isgur~\cite{Godfrey:1985xj} construct a sophisticate potential and demonstrate that the conventional quark model (CQM) or quenched quark model can explain a wide range of properties of hadrons, from the spectra to the decay widths, light mesons to heavy quarkonia.
This relativistic model is now known as GI model and was later applied to baryons~\cite{Capstick:1986ter}.
and with the newly observed data, the parameters are also refitted for $D_{(s)}$ and $B_{(s)}$ mesons~\cite{Godfrey:2015dva, Godfrey:2016nwn}.

Although the GI model has made a big achievement, it cannot be the whole story even on the theoretical side,
since the contributions of the fluctuation caused by sea quarks are totally ignored.
Additional mechanism has to be added in order to describe the hadronic decay process.
This static picture is also challenged by the experimental measurements.
For $D_s$ system, even though the ground state, such as ${}^1S_0$ and ${}^3S_1$, are well reproduced by a quenched quark model~\cite{Godfrey:1985xj},
higher charmed-strange mesons are still not well understood.

In 2003, the discovery of $D_{s0}(2317)$~\cite{BaBar:2003oey} and $D_{s}(2460)$~\cite{CLEO:2003ggt} has stimulated a lot of discussions.
The two particles are peculiar by their unexpected low masses and narrow widths compared to the quark model predictions~\cite{Godfrey:1985xj}.
Many theoretical interpretations have been proposed to address the discrepancy, including CQM with modification to the potential of GI model~\cite{Lakhina:2006fy}, hadronic molecules and compact tetraquark states~\cite{Barnes:2003dj,Yang:2021tvc,  Browder:2003fk, Lipkin:2003zk, Bicudo:2004dx, Dmitrasinovic:2005gc}.

The challenge to CQM does not stop. Recently, a new excited $D^+_s$ meson named $D_{s0}(2590)$ was observed in $B^0 \to D^-D^+K^+\pi^-$ decay by LHCb Collaboration using a data sample corresponding to an integrated luminosity of $5.4$ $\rm fb^{-1}$ at a center-of-mass energy of 13 TeV~\cite{LHCb:2020gnv}. 
The mass, total decay width, and spin-parity were detected to be $m=2591\pm6\pm7$~MeV, $\Gamma=89\pm16\pm12$~MeV, and $J^P=0^-$, respectively. 
This state was predicted to be $D_s(2^1S_0)$ with mass $2673$~MeV~\cite{Godfrey:1985xj} by the relativistic quark model, $2646$~MeV by the screen potential model~\cite{Song:2015nia}, and $2640$~MeV by the non-relativistic quark model~\cite{Li:2010vx}.
All these predictions overestimate its mass by $50-80$~MeV.
There are also works to interpret it as the $D_s(2^1S_0)$ state with $D^*K$ component~\cite{Xie:2021dwe, Ortega:2021fem} in the coupled channel framework.
For reviews on these mesons, see Refs.\cite{Chen:2016spr, Chen:2022asf} and references therein.

Although there are numerous works on $D_s$ mesons, the coupled channel effects (CCEs) still remain to be further explored.
On the side of calculation method, simple harmonic oscillator approximation of the wave function is still widely used although accurate methods are available such as Gaussian Expansion method (GEM)~\cite{Hiyama:2003cu}. 
On the theoretical side, it is common that special attention is paid to near threshold states, such as $D_s(2317)$,
or that hadronic decay widths and spectra are often separately discussed or restrained to excited $D_s$ mesons.
This may due to the fact that UQM, such as GI model, is quite successful on low lying states.
However, as will be shown in this paper, the achievement on the low lying states of the CQM does not mean that low lying states are free from the CCEs and it is more reasonable to treat all the states in a unified way. 

Moreover, for the $D_s$ mesons, the mass gap between $D_s(1968)$ and the $DK$ threshold is less than $400$~MeV.
This value is much smaller than the case of the bottomonium ($m(B\bar{B})-m(\eta_b(1S))\approx 1160$~MeV) or even the charmonium $m(D\bar{D})-m(\eta_c(1S))\approx 760$~MeV, which is a strong hint of a sizable CCE.
Since it has been shown by Li {\it et. al.} that some effect can be absorbed into the potential in UQM~\cite{Li:2009ad},
the parameters in UQM should be refitted if the CCEs are calculated in a self-consistent approach.

A systematic research of UQM for the bottomonium has been done in Ref.\cite{Lu:2016mbb}.
However, it remains unanswered whether it is possible to explain various properties of $D_s$ mesons with UQM, where CCEs are estimated to be large.
In this work, instead of fitting the lowest spectra in the CQM and only switch on the CCEs for the near-threshold states,
we perform the fitting in a fully coupled channel approach,
where the mass shift from CCEs are consistently treated for all $D_s$ mesons.
Other ingredients of CCEs are also coherently discussed, such as hadronic decays and renormalization of wave functions.

This paper is organized as follows. 
In Section~\ref{sec:formalism}, we explain the calculation framework of coupled channel effects, the details of ${}^3P_0$ model and the quenched model.
Section~\ref{sec:results} is devoted to the analysis of the results of our coupled channel calculation with various detailed comparisons and elucidation of $D_s$ mesons.
Finally, we give a short summary of this work in Section~\ref{sec:summary}

\section{Theoretical Formalism}\label{sec:formalism}

\subsection{$^3P_0$ Model and Coupled Channel Effects}
In the coupled channel framework, the full Hamiltonian is defined as
\begin{equation} \label{eqn:full}
	H = H_A + H_{BC} + H_I,
\end{equation}
where $H_A$ is the Hamiltonian of the quark-antiquark pairs.
In this work, we adopt a nonrelativistic potential model whose detailed form is discussed in the next subsection.
$H_{BC}$ is the Hamiltonian between the meson pairs, which we called BC pairs
\begin{align}
    H_{BC} = E_{BC} =\sqrt{m_B^2 +p^2} + \sqrt{m_C^2 +p^2}.
\end{align}
$H_I$ is the term which induce the mixing between $q\bar{q}$ bare state and BC meson pair system.
In this work, we use the widely used $^3P_0$ model \cite{Micu:1968mk, LeYaouanc:1972vsx, LeYaouanc:1973ldf},
where the generated quark-antiquark pairs are assumed to share the same quantum numbers with the vacuum $J^{PC} = 0^{++}$.
This assumption results in the conclusion that the spin and orbital-angular momentum to be both $1$,
thus the spectroscopy notation $^{2S +1}L_J$ of the system reads $^3P_0$.

The quark-antiquark pair-creation operator $T^\dag$ is expressed as~\cite{Ferretti:2013faa,Ferretti:2012zz,Ferretti:2013vua}
\begin{equation}
	\label{eqn:Tdag}
	\begin{array}{rcl}
	T^{\dagger} &=& -3 \, \gamma_0^{eff} \, \int d \vec{p}_3 \, d \vec{p}_4 \, 
	\delta(\vec{p}_3 + \vec{p}_4) \, C_{34} \, F_{34} \,  
	{e}^{-r_q^2 (\vec{p}_3 - \vec{p}_4)^2/6 }\,  \\
	& & \left[ \chi_{34} \, \times \, {\cal Y}_{1}(\vec{p}_3 - \vec{p}_4) \right]^{(0)}_0 \, 
	b_3^{\dagger}(\vec{p}_3) \, d_4^{\dagger}(\vec{p}_4) ~,   
	\end{array}
\end{equation}
where $C_{34}$, $F_{34}$ and $\chi_{34}$ are the color singlet wave function, flavor singlet wave function and spin triplet wave function of the $q\bar{q}$ respectively. 
$ b_3^{\dagger}(\vec{p}_3)$ and $d_4^{\dagger}(\vec{p}_4)$ are the creation operators for a quark and an antiquark with momenta $\vec{p}_3$ and $\vec{p}_4$, respectively.  
$\gamma_0^{eff}=\frac{m_n}{m_i}\gamma_0$ is the pair-creation strength, 
whose value is obtained by fitting the strong decay of the  $D_{s2}^*(2573)$ $(1^3P_2)$.
$m_n$ refers to the light quark mass $m_u$, and $m_i$ refers to the quark mass $m_u,m_d$ or $m_s$. 
In the $^3P_0$ model, the operator $T^{\dagger}$ creates a pair of constituent quarks with an actual size, 
the pair-creation point has to be smeared out by a Gaussian factor, whose width $r_q$ was determined from meson decays to be in the range $0.25$ – $0.35$ fm \cite{Silvestre-Brac:1991qqx,Geiger:1991ab,Geiger:1991qe,Geiger:1996re}. 
In our calculation, we take the value $r_q = 0.3$ fm. 

The eigenfunction of the full Hamiltonian $H$ can be expressed as,
\begin{align}
    |\psi\rangle = c_0 |\psi_0\rangle + \sum_{BC} \int d^3p\, c_{BC}(p) |BC;p\rangle,
\end{align}
where $c_0$ is the normalization constant before the $q\bar{q}$ bare state.
$c_{BC}(p)$ is the normalization constant with specific momentum $p$ for $BC$ molecular components.

Meanwhile, the eigenvalue $M$ of $H$ in Eq.~\ref{eqn:full} is the theoretical prediction of the coupled channel model which can be decomposed into two terms \cite{Kalashnikova:2005ui},
\begin{align}
\label{m}
M &= M_0 + \Delta M \\
\Delta M &= \sum_{BC} \int_0^{\infty} p^2 dp \frac{\left|\left\langle BC;p \right| T^\dagger \left| \psi_0 \right\rangle \right|^2}{M - E_{BC} + i\epsilon}
\end{align}
where $M_0$ is the eigenvalue of the quenched Hamiltonian $H_A$ and $\Delta M$ is the mass shift which signifies the deviation between $H$ and $H_A$.
If the initial state $A$ is above the threshold of $B$ and $C$, 
a strong decay process $A\to BC$ will happen and $\Delta M$ will pick up a imaginary part which equals to the one half of the decay width.
\begin{equation}
\label{eqn:decay}
    \Gamma_{BC} = 2 \pi p_0 \frac{E_B(p_0) E_C(p_0)}{m_A} \left| \left\langle BC;p_0\right| T^\dagger \left| \psi_0 \right\rangle \right|^2
\end{equation}

For states below the $BC$ threshold, it is possible to normalize the physical state $|\psi\rangle$, 
and the probability of quenched quark pairs can be calculated as,
\begin{equation}
\label{eqn:pqqbar}
	P_{q\bar{q}} := |c_0|^2 = \left(1+\sum_{BC} \int_0^{\infty} p^2 dp \frac{\left|\left\langle BC;p \ell J \right| T^\dagger \left| \psi_0 \right\rangle \right|^2}{(M - E_{BC})^2}\right)^{-1},
\end{equation}
and the probability of the molecular component are naturally expressed as $P_{BC}= 1- P_{q\bar{q}}$.
If the states locate above the threshold, it is not possible to normalize the wave functions.
However, we can still estimate the proportion of the closed channel by 
\begin{equation}
\label{eqn:pbc}
    P_{BC} := \int d^3p \, |c_{BC(p)}|^2 = \int_0^{\infty} p^2 dp \frac{\left|\left\langle BC;p \right| T^\dagger \left| \psi_0 \right\rangle \right|^2}{(M - E_{BC})^2},
\end{equation}
which is equivalent to fix $P_{q\bar{q}}=1$.
When the probabilities is expressed in this form, it is convenient to estimate and compare the proportion of different components.
In this work, we adopt the assumption that the opened channels contribution to the wave functions can be discarded \cite{Heikkila:1983wd},
which will make the normalization proceed.

As can be seen from above, the occurrence of $H_I$ will not only result in a mass shift $\Delta M$ to the spectrum and a decay process but also renormalize the wave function.
These three effects are jointly called coupled channel effects (CCEs) which we aim to explore in the next section.
For other aspects of the CCEs like $S-D$ mixing, we encourage the readers to Ref.\cite{Lu:2016mbb}.

\subsection{Non-relativistic quark model}
The quenched part of the Hamiltonian $H_A$ is taken from the a potential model where $\alpha_s^2$ correction is explicitly addressed.
This model was proposed by Lakhina and Swanson~\cite{Lakhina:2006fy}, and has been used to study the bottom mesons~\cite{Lu:2016bbk} and open charm mesons~\cite{Li:2010vx}. 
The Hamiltonian $H_A$ can be split into a term $H_0$ which could be solved non-perturbatively and a spin-dependent term $H_{sd}$ which we solve in leading order perturbation theory.
\begin{align}
    H_A =& H_0+H_{sd},\\
    H_0 =& m_{q} + m_{\bar{q}} + \frac{\nabla^2}{2M_r}-\frac{4}{3}\frac{{\alpha}_s}{r}+br+C_{q\bar{q}}\nonumber\\  & + \frac{32{\alpha}_s{\sigma}^3 e^{-{\sigma}^2r^2}}{9\sqrt{\pi}m_qm_{\bar{q}}} {\boldsymbol{S}}_{q} \cdot {\boldsymbol{S}}_{\bar{q}},
\end{align}
where $M_r$ is the reduced mass of which equals to $M_r=m_q m_{\bar{q}}/(m_q+m_{\bar{q}})$. 
When combined together with the subscript, ${\boldsymbol{S}}$ and $m$ stand for the spin and mass of quark or antiquark, respectively.
where $m_{q}, m_{\bar{q}}, \alpha_s, b, C_{q\bar{q}}$ and $\sigma$ are the free parameters we need to refit in this quenched model.

The $H_{sd}$ can be compressed as
    \begin{eqnarray}
      H_{sd} &=& \left(\frac{\boldsymbol{S}_{q}}{2m_q^2}+\frac{{\boldsymbol{S}}_{\bar{q}}}{2m_{\bar{q}}^2}\right) \cdot \boldsymbol{L}\left(\frac{1}{r}\frac{dV_c}{dr}+\frac{2}{r}\frac{dV_1}{dr}\right)\nonumber\\
      &&+\frac{{\boldsymbol{S}}_+ \cdot \boldsymbol{L}}{m_qm_{\bar{q}}}\left(\frac{1}{r} \frac{dV_2}{r}\right) \nonumber\\
      && +\frac{3{\boldsymbol{S}}_{q} \cdot \hat{\boldsymbol{r}}{\boldsymbol{S}}_{\bar{q}} \cdot \hat{\boldsymbol{r}}-{\boldsymbol{S}}_{q} \cdot {\boldsymbol{S}}_{\bar{q}}}{3m_qm_{\bar{q}}}V_3\nonumber\\
      && +\left[\left(\frac{{\boldsymbol{S}}_{q}}{m_q^2}-\frac{{\boldsymbol{S}}_{\bar{q}}}{m_{\bar{q}}^2}\right)+\frac{{\boldsymbol{S}}_-}{m_qm_{\bar{q}}}\right] \cdot \boldsymbol{L} V_4,
\end{eqnarray}
where
\begin{eqnarray}
  V_c &=& -\frac{4}{3}\frac{{\alpha}_s}{r}+br,\nonumber \\
  V_1 &=& -br-\frac{2}{9\pi}\frac{{\alpha}_s^2}{r}[9{\rm ln}(\sqrt{m_qm_{\bar{q}}}r)+9{\gamma}_E-4],\nonumber\\
  V_2 &=& -\frac{4}{3}\frac{{\alpha}_s}{r}-\frac{1}{9\pi}\frac{{\alpha}_s^2}{r}[-18{\rm ln}(\sqrt{m_qm_{\bar{q}}}r)+54{\rm ln}(\mu r)\nonumber\\
  &&+36{\gamma}_E+29],\nonumber\\
  V_3 &=& -\frac{4{\alpha}_s}{r^3}-\frac{1}{3\pi}\frac{{\alpha}_s^2}{r^3}[-36{\rm ln}(\sqrt{m_qm_{\bar{q}}}r)+54{\rm ln}(\mu r)\nonumber\\
  &&+18{\gamma}_E+31],\nonumber\\
  V_4 &=& \frac{1}{\pi}\frac{{\alpha}_s^2}{r^3}{\rm ln}\left(\frac{m_{\bar{q}}}{m_q}\right),
\end{eqnarray}
where $\boldsymbol{S}_{\pm}={\boldsymbol{S}}_q\pm{\boldsymbol{S}}_{\bar{q}}$, $\boldsymbol{L}$ is the orbital angular momentum of the $q\bar{q}$ system. 
$\gamma_E$ is Euler constant, $C_F$ and $C_A$ are gauge group factors and $\mu$ is renormalization scale. 
The value of the parameters are taken from Ref.\cite{Lakhina:2006fy}: $\gamma_E=0.5772$, $C_F=4/3$, $C_A=3$ and $\mu=1$~GeV.

The spin-orbit term in the $H_{sd}$ can be rewritten into
symmetric part $H_{sym}$ and antisymmetric part $H_{anti}$,
\begin{eqnarray}
H_{sym} &=& \frac{{\boldsymbol{S}}_+ \cdot {\boldsymbol{L}}}{2}\left[\left(\frac{1}{2m_q^2}+\frac{1}{2m_{\bar{q}}^2}\right) \left(\frac{1}{r}\frac{dV_c}{dr}+\frac{2}{r}\frac{dV_1}{dr}\right)\right. \nonumber \\
&& \left.+\frac{2}{m_qm_{\bar{q}}}\left(\frac{1}{r} \frac{dV_2}{r}\right)+\left(\frac{1}{m_q^2}-\frac{1}{m_{\bar{q}}^2}\right)V_4\right],
\end{eqnarray}
\begin{eqnarray}
H_{anti} &=& \frac{{\boldsymbol{S}}_- \cdot {\boldsymbol{L}}}{2}\left[\left(\frac{1}{2m_q^2}-\frac{1}{2m_{\bar{q}}^2}\right) \left(\frac{1}{r}\frac{dV_c}{dr}+\frac{2}{r}\frac{dV_1}{dr}\right)\right. \nonumber \\
&& \left.+\left(\frac{1}{m_q^2}+\frac{1}{m_{\bar{q}}^2}+\frac{2}{m_qm_{\bar{q}}}\right)V_4\right].
\end{eqnarray}
The antisymmetric part $H_{anti}$ gives rise to the
the spin-orbit mixing of the mesons, such as $D_s(n{}^3L_L)-D_s(n{}^1L_L)$. 
This mixing can be parametrized by a mixing angle $\theta_{nL}$ by the following formula \cite{Godfrey:1985xj,Godfrey:1986wj},
\begin{equation}
\left(
\begin{array}{cr}
D_{sL}(nL)\\
D^\prime_{sL}(nL)
\end{array}
\right)
 =\left(
 \begin{array}{cr}
\cos \theta_{nL} & \sin \theta_{nL} \\
-\sin \theta_{nL} & \cos \theta_{nL}
\end{array}
\right)
\left(\begin{array}{cr}
D_s(n^1L_L)\\
D_s(n^3L_L)
\end{array}
\right),
\label{eqn:mix}
\end{equation}
where $D_{sL}(nL)$ and $D_{sL}^\prime(nL)$ represent the physical observed states.

\section{Results and discussions}
\label{sec:results}

After solving the Schr\"{o}dinger equation with Hamiltonian $H_0$, 
the wave functions are used to determine the leading order correction of $H_{sd}$ in the quenched quark model and the CCEs to the spectrum and strong decays and wave functions.

The refitted parameters are listed in Table~\ref{tab:para}.
They are fixed by fitting the strong decay of the $D_{s2}^*(2573)$ and the spectrum of 
$D_s(1^1S_0)$, $D_s^*(1^3S_1)$, $D_{s0}^*(2317)(1^3P_0)$, $D_{s2}^*(2573)(1^3P_2)$, $D_{s3}^*(2860)(1^3D_3)$, $D_{s1}^*(2590)(2^1S_0)$ and $D_{s1}^*(2700)(2^3S_1)$, and all the input values are taken from PDG~\cite{PDG:2021}. We also list these parameters from the previous quark model. That indicates that the coupled channel effects can cause the difference.

\begin{figure}
    \centering
    \includegraphics[width=\columnwidth]{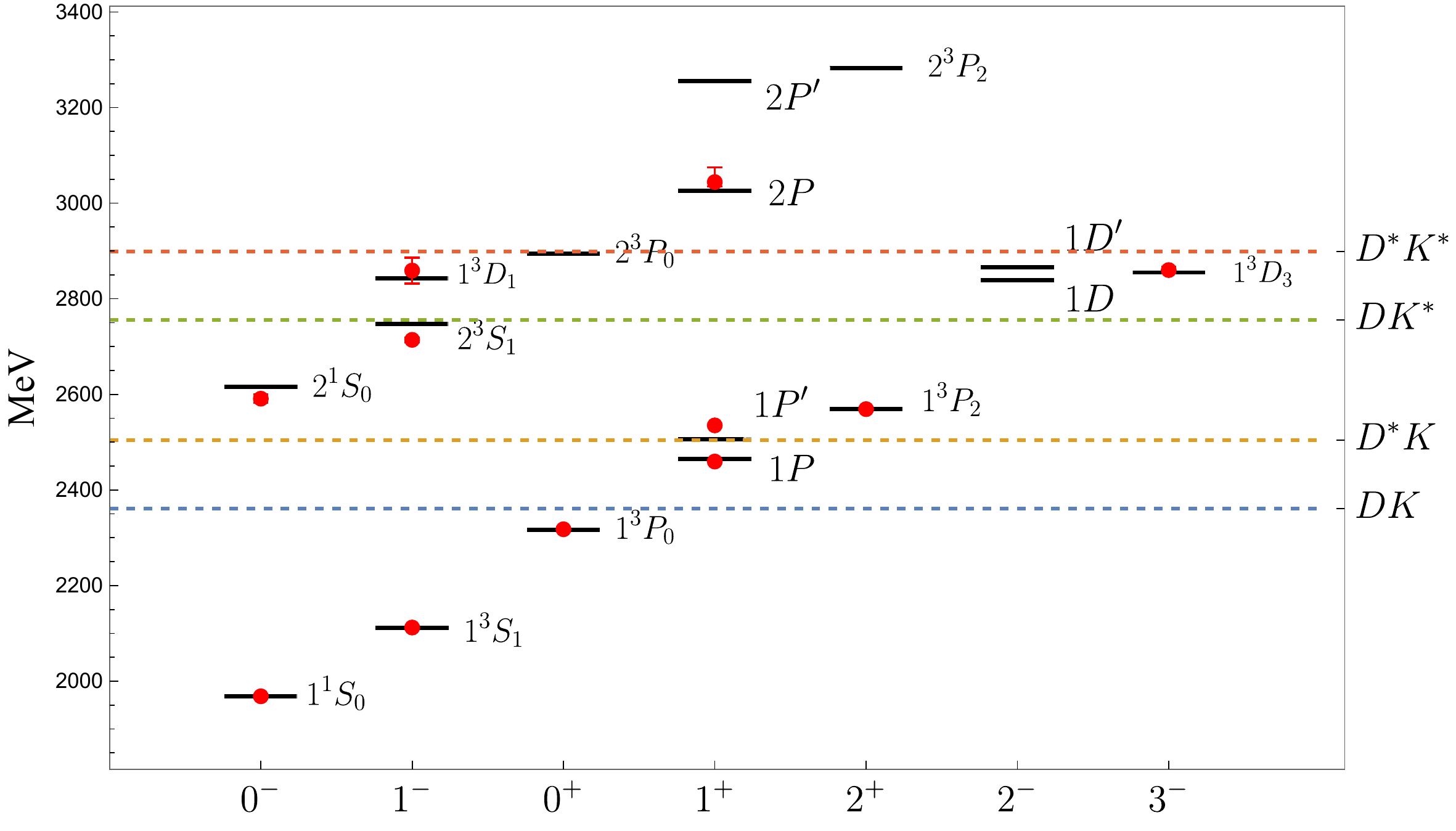}
    \caption{The spectrum of the $D_s$ mesons. Red dots with error bars denote the experimental values from PDG~\cite{PDG:2021} and our calculations are depicted as black lines.}
    \label{fig:spectrum}
\end{figure}

\begin{table}[h] 
\caption{Parameters refitted in this work.} 
\label{tab:para}
\begin{center}
\begin{tabular}{ccc} 
\hline 
\hline \\
Parameter  &  This work            &Ref. \cite{Li:2010vx} \\ \\
\hline \\
$m_n$      & $0.419$ GeV      & $0.45$ GeV \\
$m_s$      & $0.569$ GeV      & $0.55$ GeV\\
$m_c$      & $1.464$ GeV      & $1.43$ GeV\\   \\
$\alpha_s$ & $0.6072$         & $0.5$  \\  
$b$        & $0.1314$ GeV$^2$ & $0.14$ GeV$^2$\\  
$\sigma$   & $1.2071$ GeV     & $1.17$ GeV \\
$C_{cs}$   & $0.1381$ GeV     & $-0.325$ GeV \\  \\
$\gamma_0$ & $0.529$          & $0.452$  \\
\hline 
\hline
\end{tabular}
\end{center}
\end{table}
With the above fitting parameters, we get the following mixing angles in Eq.\eqref{eqn:mix} to be 
$-29.5^\circ$, $-32.7^\circ$ and $-41.9^\circ$ for $1P$, $2P$ and $1D$, respectively. In the screened potential model the mixing angle could be $-42.7^\circ$, $-31.4^\circ$ and $-39.4^\circ$ \cite{Gao:2022bsb}. In the quenched quark model it could be $-24.5^\circ$, $-32.3^\circ$ and $-40.2^\circ$ \cite{Li:2010vx}. Although the exact numbers for the mixing angles in various quark models are different, they are close.
Results on the spectrum are shown in Fig.\ref{fig:spectrum} with numbers listed in Table~\ref{tab:spectrum}.

\begin{table*}[htpb]
\begin{center}
\caption{\label{tab:decay1} The decay widths of the $2S$ and $1P$ states respectively. $-$ means the channel is forbidden.}
\footnotesize
\begin{tabular}{lcccccc}
\hline\hline

  Channel            & $2^1S_0$        &$2^3S_1$          &$1^3P_0$     &$1P$     &$1P^\prime$           &$1^3P_2$    \\   
                     & $D_{s0}(2590)$  &$D_{s1}^*(2700)$  & $D_{s0}^*(2317)$      & $D_{s1}(2460)$       & $D_{s1}(2536)$  & $D_{s2}^*(2573)$   \\\hline
  $DK$               &$-$              &22                &$-$          &$-$      &$-$                   &15          \\
  $DK^*$             &$-$              &$-$               &$-$          &$-$      &$-$                   &$-$          \\
  $D^*K$             &112              &92                &$-$          &$-$      &0.4                     &2           \\
  Total              &112              &114               &$-$          &$-$      &0.4                     &17          \\
  Exp.               &$89\pm16\pm12$\cite{LHCb:2020gnv}  &$122\pm10$  &$<3.8$  &$<3.5$  &$0.92\pm0.05$   &$16.9\pm0.7$  \\
  \hline\hline

\end{tabular}
\end{center}
\end{table*}

\begin{table*}[htpb]
\begin{center}
\caption{\label{tab:decay2} The decay width of the $2P$ and $1D$ states respectively. $-$ means the channel is forbidden or no experimental information.}
\footnotesize
\begin{tabular}{lccccccccc}
\hline\hline

  Channel               &$2^3P_0$   &$2P$              &$2P^\prime$   &$2^3P_2$    &$1^3D_1$           &$1D$     &$1D^\prime$    &$1^3D_3$\\   
                        &$-$        & $D_{sJ}^*(3040)$ &$-$           &$-$         & $D_{s1}^*(2860)$  &$-$      &$-$            &$D_{s3}^*(2860)$ \\\hline
  $DK$                  &41         &$-$                &$-$           &20          &31                 &$-$      &$-$            &32  \\
  $DK^*$                &$-$        &52                &12            &1           &35                 &99       &11             &3  \\
  $D^*K$                &$-$        &12                 &8             &13          &28                 &45       &78             &29 \\
  $D^*K^*$              &$-$        &166               &197           &200         &$-$                &$-$      &$-$            &$-$\\
  $D_s\eta$             &5          &$-$               &$-$           &1           &9                  &$-$      &$-$            &2  \\
  $D_s\eta^\prime$      &$-$        &$-$               &$-$           &0.4         &$-$                &$-$      &$-$            &$-$ \\
  $D_s\phi$             &$-$        &18                &21            &12          & $-$               &$-$      &$-$            &$-$ \\
  $D_s^*\eta$           &$-$        &6                 &7             &0.02        &5                  &1        &17             &1   \\
  $D_s^*\eta^\prime$    &$-$        &$-$               &1             &2           &$-$                &$-$      &$-$            &$-$ \\
  $D_s^*\phi$           &$-$        &$-$               &41            &31          &$-$                &$-$      &$-$            &$-$   \\
  $DK^*_0(1430)$        &$-$        &$-$               &$-$           &$-$         &$-$                &$-$      &$-$            &$-$   \\
  $DK_{1B}$             &$-$        &$-$               &5             &20          &$-$                &$-$      &$-$            &$-$   \\
  $DK_{1A}$             &$-$        &$-$               &$-$           &0.2         &$-$                &$-$      &$-$            &$-$   \\
  $DK^*_2(1430)$        &$-$        &$-$               &$-$           &$-$         &$-$                &$-$      &$-$            &$-$   \\
  $D_0^*(2300)K$        &$-$        &3                 &0.2           &$-$         &$-$                &$-$      &0.01           &$-$   \\
  $D_1(2420)K$          &$-$        &16                &3             &66          &$-$                &$-$      &$-$            &$-$   \\
  $D_1(2430)K$          &$-$        &0.03              &0.005          &9           &$-$                &$-$      &$-$            &$-$   \\
  $D_2^*(2460)K$        &$-$        &5                 &38            &46          &$-$                &$-$      &$-$            &$-$   \\
  Total                 &47         &279               &331           &421         &110                &145      &106            &67  \\
  Exp.                  & $-$       &$239\pm60$        &$-$           & $-$       &$159\pm80$          &$-$      &$-$            &$53\pm10$ \\
  \hline\hline

\end{tabular}
\end{center}
\end{table*}

\begin{table*}[htpb]
\begin{center}
\caption{\label{tab:spectrum} The mass spectrum (in MeV) of the $c\bar{s}$ mesons.
Column 3 to 5 stand for spectrum from the potential model, the mass shift, the spectrum with coupled channel effects, respectively. 
Results from Ref.~\cite{Li:2010vx} is listed in Column 6 as comparison.
The last Column is the experimental values taken from PDG \cite{PDG:2021}.
}
\footnotesize
\begin{tabular}{ccccccc}
\hline\hline
  $n^{2S+1}L_J$  & state             &$M_0$  &$\Delta M$  &$M$  &NR \cite{Li:2010vx}   & PDG~\cite{PDG:2021}  \\\hline
  $1^1S_0$     & $D_{s}$              &$2272$   &$-304$     &$1968$   &$1969$   &$1968.34\pm0.07$    \\
  $1^3S_1$     & $D_{s}^{*}$          &$2472$   &$-359$     &$2112$   &$2107$   &$2112.2\pm0.4$    \\
  $2^1S_0$     & $D_{s0}(2590)$       &$2989$   &$-373$     &$2616$   &$2640$   &$2591\pm6\pm7$\cite{LHCb:2020gnv}           \\
  $2^3S_1$     & $D_{s1}^*(2700)$     &$3081$   &$-334$    &$2747$   &$2714$   &$2714\pm5$                      \\
  $1^3P_0$     & $D_{s0}^*(2317)$     &$2668$   &$-351$     &$2316$   &$2344$   &$2317.8\pm0.5$      \\
  $1P$         & $D_{s1}(2460)$       &$2843$   &$-378$     &$2465$   &$2488$   &$2459.5\pm0.6$       \\
  $1P^\prime$  & $D_{s1}(2536)$       &$2897$   &$-392$     &$2506$   &$2510$   &$2535.11\pm0.06$     \\
  $1^3P_2$     & $D_{s2}^*(2573)$     &$2959$   &$-390$     &$2569$   &$2559$   &$2569.1\pm0.8$     \\
  $2^3P_0$     & $-$                   &$3151$   &$-257$     &$2894$   &$2830$   &$-$ \\
  $2P$         & $D_{sJ}^*(3040)$      &$3302$   &$-276$     &$3026$   &$2958$   &$3044^{+31}_{-9}$  \\
  $2P^\prime$  &  $-$                   &$3367$   &$-112$     &$3255$   &$2995$   &$-$   \\
  $2^3P_2$     & $-$                   &$3422$   &$-138$     &$3283$   &$3040$   &$-$ \\
  $1^3D_1$     & $D_{s1}^*(2860)$     &$3159$   &$-315$     &$2843$   &$2804$   &$2859\pm27$    \\
  $1D$         & $-$                   &$3175$   &$-336$     &$2839$   &$2788$   &$-$    \\
  $1D^\prime$  & $-$                   &$3216$   &$-350$     &$2866$   &$2849$   &$-$    \\
  $1^3D_3$     & $D_{s3}^*(2860)$     &$3206$   &$-350$     &$2855$   &$2811$   &$2860\pm7$       \\
  \hline\hline
\end{tabular}
\end{center}
\end{table*}

\begin{table*}
\caption{\label{tab:shift} Mass shift $\Delta M$ (in MeV) of each coupled channel.} 
\begin{tabular}{cccccccccccc} 
\hline 
\hline \\
State                &$DK$  &$DK^*$ &$D^*K$ &$D^*K^*$ &$D_s\eta$ &$D_s\eta^\prime$ &$D_s\phi$  &$D_s^*\eta$  &$D_s^*\eta^\prime$ &$D_s^*\phi$ &Total \\ \\
\hline \\
$1^1S_0$             &$0$      &$-61$    &$-43$    &$-130$  &$0$   &$0$     &$-17$    &$-10$    &$-4$  &$-39$  &$-304$  \\  
$1^3S_1$             &$-19$    &$-48$    &$-35$    &$-176$  &$-4$  &$-1$    &$-13$    &$-8$     &$-3$  &$-53$  &$-359$  \\  
$2^1S_0$             &$0.0$    &$-77$    &$-88$     &$-149$ &$0$   &$0$     &$-14$     &$-11$    &$-3$  &$-32$  &$-373$  \\
$2^3S_1$             &$2$      &$-70$    &$-13$    &$-194$  &$-2$  &$-1$    &$-9$     &$-9$     &$-2$  &$-37$  &$-334$ \\    
$1^3P_0$             &$-62$    &$0$      &$0$      &$-222$  &$-6$  &$-2$    &$0$      &$0$      &$0$  &$-59$  &$-351$  \\ 
$1P$                 &$0$      &$-56$    &$-68$    &$-183$  &$0$   &$0$     &$-13$    &$-9$     &$-3$  &$-47$  &$-378$  \\     
$1P^\prime$          &$0$      &$-75$    &$-89$    &$-156$  &$0$   &$0$    &$-17$     &$-11$   &$-4$   &$-39$  &$-392$  \\ 
$1^3P_2$             &$-37$    &$-59$    &$-49$    &$-172$  &$-5$    &$-2$   &$-13$     &$-8$   &$-3$  &$-42$   &$-390$  \\
$2^3P_0$             &$-3$     &$0$      &$0$      &$-217$  &$-3$  &$-1$     &$0$      &$0$      &$0$  &$-33$  &$-257$  \\ 
$2P$                 &$0$      &$-33$    &$-8$    &$-195$  &$0$    &$0$    &$-7$     &$-4$   &$-2$   &$-27$  &$-276$  \\     
$2P^\prime$          &$0$      &$-16$    &$-28$    &$-20$  &$0$    &$0$    &$-8$     &$-3$   &$-3$   &$-34$  &$-112$  \\  
$2^3P_2$             &$-19$    &$-22$    &$-30$    &$-23$  &$-3$    &$-1$   &$-4$     &$-3$   &$-2$  &$-32$   &$-138$  \\ 
$1^3D_1$             &$13$     &$-19$    &$4$      &$-257$   &$0.3$   &$-1$   &$-2$     &$-2$   &$-1$   &$-52$  &$-315$  \\
$1D$                 &$0$      &$-75$    &$-24$    &$-180$  &$0$    &$0$    &$-11$     &$-8$   &$-2$   &$-36$  &$-336$  \\ 
$1D^\prime$          &$0$      &$-83$    &$-41$    &$-168$  &$0$    &$0$    &$-14$     &$-9$   &$-3$   &$-31$  &$-350$  \\
$1^3D_3$             &$-28$    &$-60$    &$-45$    &$-164$  &$-5$    &$-1$   &$-10$     &$-7$   &$-2$   &$-29$  &$-350$ \\ 
\hline 
\hline
\end{tabular}
\end{table*}

\begin{table*}
\caption{\label{tab:probabilities} Probabilities (in $\%$) of the coupled channels considered in this work. 
For the convenience of comparison, values from Column 3 to 12 (various coupled channels) are rescaled by $c\bar{s}$ values, such that $P_{c\bar{s}}=100\%$.
``-'' stands for that the corresponding channel is open and its contribution to the wave function normalization is discarded, see discussion in the Section. 
Last two columns represent the probability of molecular components and $c\bar{s}$, respectively.}
\begin{tabular}{ccccccccccccccc} 
\hline 
\hline \\
$(n_r+1){}^{2S+1}L_J$   &State  &$DK$   &$DK^*$ &$D^*K$ &$D^*K^*$   &$D_s\eta$  &$D_s\eta^\prime$ &$D_s\phi$ &$D_s^*\eta$  &$D_s^*\eta^\prime$ &$D_s^*\phi$ &$P_{\mathrm{molecule}}$  & $P_{c\bar{s}}$\\ \\
\hline \\
$1^1S_0$    &$D_s$             &$0.0$  &$4.3$  &$3.5$  &$8.5$      &$0.0$     &$0.0$            &$1.1$     &$0.7$       &$0.2$  &$2.2$  &$17.0$  &$83.0$ \\  
$1^3S_1$    &$D_{s}^{*}$       &$2.5$  &$4.2$  &$3.8$  &$13.9$     &$0.4$     &$0.1$            &$1.0$     &$0.7$       &$0.2$  &$3.5$  &$23.2$  &$76.8$ \\  
$1^3P_0$    &$D_{s0}^*(2317)$  &$45.5$ &$0.0$  &$0.0$  &$19.9$     &$1.7$     &$0.2$            &$0.0$     &$0.0$       &$0.0$  &$4.2$  &$40.3$  &$59.7$ \\ 
$1P$        &$D_{s1}(2460)$    &$0.0$  &$8.5$  &$42.8$ &$19.1$     &$0.0$     &$0.0$            &$1.3$     &$1.8$       &$0.3$  &$3.8$  &$43.7$  &$56.3$ \\
\hline
$1P^\prime$  & $D_{s1}(2536)$  &$-$    &$10.8$ &$-$    &$17.9$     &$-$       &$-$              &$1.7$     &$1.9$       &$0.4$  &$3.4$  &$26.5$  &$73.5$ \\  
$1^3P_2$     & $D_{s2}^*(2573)$&$-$    &$8.5$  &$-$    &$22.8$     &$-$       &$0.2$            &$1.4$     &$1.2$       &$0.3$  &$4.0$  &$27.7$  &$72.3$ \\
$2^1S_0$     & $D_{s0}(2590)$  &$-$    &$20.4$ &$-$    &$26.2$     &$-$       &$-$              &$2.0$     &$4.1$       &$0.4$  &$3.7$  &$36.2$  &$63.8$ \\
$2^3S_1$     & $D_{s1}^*(2700)$&$-$    &$51.3$ &$-$    &$47.3$     &$-$       &$0.2$            &$1.6$     &$-$         &$0.3$  &$4.7$  &$51.3$  &$48.7$ \\
$1^3D_1$     & $D_{s1}^*(2860)$&$-$    &$-$    &$-$    &$47.6$     &$-$       &$0.5$            &$0.6$     &$-$         &$0.1$  &$5.8$  &$35.3$  &$64.7$ \\
$1D$         & $-$             &$-$    &$-$    &$-$    &$35.4$     &$-$       &$-$              &$2.0$     &$-$         &$0.4$  &$4.1$  &$29.5$  &$70.5$ \\
$1D^\prime$  & $-$             &$-$    &$-$    &$-$    &$46.9$     &$-$       &$-$              &$2.3$     &$-$         &$0.4$  &$3.9$  &$34.9$  &$65.1$ \\
$1^3D_3$     & $D_{s3}^*(2860)$&$-$    &$-$    &$-$    &$54.4$     &$-$       &$0.2$            &$1.4$     &$-$         &$0.3$  &$3.8$  &$37.5$  &$62.5$ \\
$2^3P_0$     & $-$             &$-$    &$-$    &$-$    &$167.5$    &$-$       &$0.6$            &$-$       &$-$         &$-$    &$4.0$  &$63.2$  &$36.8$ \\   
\hline 
\hline
\end{tabular}
\end{table*}

A general conclusion is that all the $D_s$ mesons contains sizable molecular components, among which, $D^*K^*$ contributes dominantly.
This large contribution is mainly due to the spin enhancement effect.

For states below $DK$ thresholds, the wave function of $D_s$ mesons can be normalized, 
and our calculation shows that even the ground state $1{}^1S_0$, which is widely accepted as a pure $c\bar{s}$ in the quenched picture, contains $17\%$ molecular components.
As we have explained in the introduction, the $DK$ threshold is much lower for $D_s$ mesons, and this result verifies what we have anticipated:
the CCEs are important for the study of the $D_s$ mesons, even for the ground states.

We need to point out that even though the CQM have the ability to reproduce the spectrum, this does not mean that CQM is an good approximation for the ground state $D_s$ mesons. 
The achievement of CQM on ground state is mainly due to the fact that the CCEs can be partially absorbed into the parameters. e.g., the constant term in the potential can be leveraged to absorb the mass shift in CCEs.
However, as we have discussed in the previous section, the mass shift only reflects one aspect of the CCEs. 
For other impacts in CCEs, like the renormalization of the wave functions, there is no way to embrace it in the CQM since it is conceptually beyond the CQM.

The breakdown of the CQM can also be revealed by the famous $D_{s0}^*(2317)$ and $D_{s1}^*(2460)$.
The CQM like GI model \cite{Godfrey:2015dva} predict their masses to be $167$ MeV and $89$MeV higher, respectively.
As a comparison, with CCEs consistently calculated, our results agrees excellently with the PDG results.
Since the decay channel $DK$ above $D_{s0}^*(2317)$ and as a $J^P =1^+$ particle, $D_{s1}^*(2460)$ cannot decay to two $0^-$ final state.
The only allowed hadronic decay is the isospin symmetry breaking process $D_s^* \to D_s \pi$, resulting a narrow width for both.

Another impact of the Table~\ref{tab:probabilities} is that $40.3\%$ of $D_{s0}^*(2317)$ is make up of various molecular components and the dominant $DK$ molecule is $27.2\%$.
$D_{s1}(2460)$ contains a larger molecular componets ($43.7\%$) where dominent component becomes $D^*K$ ($24.1\%$).
The large molecular components also offers a partial reason why it is difficult to incorporate them in the quenched quark model.

The mixing between $D_{s1}(2460)$ and $D_{s1}(2536)$ are assumed to be caused by the anti-symmetric part in the spin-orbit Hamiltonian $H_{anti}$,
and with our fitted parameters, the mixing angle is predicted to be $-29.5^\circ$.
As can be seen from the Table~\ref{tab:spectrum}, 
our prediction of $D_{s1}(2460)$ is only 5 MeV above the experimental central value.

A noticeable deviation is the mass of $D_{s1}(2536)$, our prediction is about $30$~MeV lighter than the PDG averaged.
However, this result could be improved  within the coupled channel framework, 
since additional off-diagonal term which represents the mixing between bare state and the molecular state will enlarge the mass gap between $1P$ and $1P'$, thus lifting the mass of $1P'$ to be closer to the experimental measures.
This requires a refitting of all the parameters and we postpone this challenge for later work.

The probabilities of each coupled channel are listed in column 3 - 12 in Table.~\ref{tab:probabilities}, where $P_{c\bar{s}}$ is set to 1 for comparison convenience.
One general feature is that $P_{\mathrm{molecule}}$ is not negligible even for the ground state $D_s$ mesons. 
This may explain why it is a challenge to fit the spectrum in CQM because of the universal CCEs.

For states above threshold, the additional quark pairs leads to hadronic decay.
As can be seen from the Table~\ref{tab:decay1} and \ref{tab:decay2},
our prediction on the hadronic decay widths agrees excellently with the experimental measurement,
given that our fit focuses mainly on the spectrum and the only hadronic decay with in the fitting is that of $D_{s2}^*(2573)$.

For $D_{s1}^*(2700)$, our result $2747$ MeV is more close to the newly LHCb measurement $2732.3\pm4.3\pm5.8$ MeV \cite{LHCb:2016mbk} instead of the PDG averaged $2714\pm 5$ MeV.
And its molecular components could be as large as $51.3\%$.
This also challenges the assignment that $D_{s1}^*(2700)$ is a good candidate of pure $2{}^3S_1$.
This result can be further improved when ${}^3S_1-{}^3D_1$ mixing introduced by CCEs is considered.
Additionally, because a cross-term in the Hamiltonian will increase the mass splitting, this $S-D$ mixing also have the potential to improve our result of $D_{s1}(2860)$.
Since the mass uncertainties of $D_{s1}(2860)$ is still relatively large, and $S-D$ mixing is not the center of this work, we postpone the study of the fine structure to later work.

As the name suggest, $D_{s1}(2860)$ and $D_{s3}(2860)$ are nearly equally heavy with each other.
Although the GI model also predicts similar mass splitting, $2899$ and $2917$ MeV for $D_{s1}({}^3D_1)$ and $D_{s1}({}^3D_3)$, respectively \cite{Godfrey:2015dva}.
The experimental results are $40\sim60$ MeV lighter than what the GI model has claimed.
As a comparison, our results reproduce the spectra very well.
Furthermore, our calculation not only reproduces the total decay widths of both particles,  
but also matches the branch ratio of $\Gamma(D_s(2860) \to D^*K)/\Gamma(D_s(2860) \to DK)$.
As can be read from the Table~\ref{tab:decay2}, the ratios are around 0.9 for both $D_{s1}(2860)$ and $D_{s3}(2860)$, which agrees with the measurement from BaBar \cite{BaBar:2009rro}  ($1.10 \pm 0.15 \pm 0.19$). 

For $D_s(3040)$, the total decay width also matches the experimental measured values \cite{BaBar:2009rro} $239\pm 35 ^{+46}_{-42}$~MeV and our calculation suggests that the main decay channel is $D^*K^*$.
Since the main decay channel of $K^*$ is $K^*\to K \pi$, we suggest to search at the channel $D^*K\pi$ to verify our prediction.

One important prediction in this work is about the property of $D_s(2{}^3P_0)$. 
Our calculation shows that its mass is 2894 MeV, only 5~MeV below $D^*K^*$ threshold. 
This value is 111~MeV below the GI Model prediction~\cite{Godfrey:2015dva}.
The hadronic width is 47~MeV and it decays mainly to $DK$.
It also couples strongly to $D^*K^*$, and as suggested in Table~\ref{tab:probabilities}, this coupling is around 1.7 as large as that of the $c\bar{s}$ core.
This exceedingly large coupling concludes that around $62\%$ of $2{}^3P_0$ consists of $D^*K^*$ molecule,
which is quite beyond the conventional quark model, and we suggest to search for related signals experimentally.

\section{Summary}
\label{sec:summary}

We made a coupled channel calculation of the $D_s$ mesons, where the spectrum and the decay width are coherently calculated.
The decent match with the experiment on the spectra, decay widths or even the decay branch ratios givess us a strong evidence that the coupled channel effects are able to explain both the ground and the excited $D_s$ mesons.
Some exotic states like $D_s(2317)$ can also be nicely explained in this framework.
We also get universally sizable non-$c\bar{s}$ components for all the $D_s$ mesons, which also signifies a strong coupled channel effects.
One of our various predictions is on the property of $D_s(2{}^3P_0)$, we claim that it should be a $D^*K^*$ dominant molecule with mass $2894$~MeV.
Its hadronic decay width is $47$~MeV with dominant decay channel $DK$.

\section{Acknowledgements}
The authors are grateful to Feng-kun Guo and Jia-jun Wu for valuable suggestions and comments.
This work is supported by the NSFC and the Deutsche Forschungsgemeinschaft (DFG, German Research Foundation) through the funds provided to the Sino-German Collaborative Research Center TRR110 Symmetries and the Emergence of Structure in QCD (NSFC Grant No. 12070131001, DFG Project-ID 196253076-TRR 110), by the NSFC Grants No. 11835015 and No. 12047503, and by the Chinese Academy of Sciences (CAS) under Grant No. XDB34030000.

\bibliography{cite}  

\end{document}